\documentclass[aps,prb,floats,floatfix,twocolumn]{revtex4}
 
\usepackage{ifthen}
\usepackage{ifpdf}
\usepackage{color}
\ifpdf
\usepackage{graphicx}
\usepackage{epstopdf}
\else
\usepackage{graphicx}
\usepackage{epsfig}
\fi
\graphicspath{{./Figs/}{./}}
\usepackage{ulem}
\usepackage{latexsym}
\usepackage{amsmath}
\usepackage{amssymb}
\usepackage{bm}
\usepackage{wasysym}

\usepackage{latexsym}
\usepackage{amsmath}
\usepackage{amssymb}
\usepackage{bm}
\usepackage{wasysym}
\usepackage{ulem}

\begin{document}
\title{Quantum signatures of chaos in a cavity-QED-based stimulated Raman adiabatic passage }

\author{Amit  Dey\footnote{amit.dey.85@gmail.com}}
\affiliation{Ramananda College, Bankura University, Bankura 722122, India\\}
\affiliation{International Centre for Theoretical Sciences, Tata Institute of Fundamental Research, Bengaluru -- 560089, India}

\date{\today}
\begin{abstract}
 \noindent Nonlinear stimulated Raman adiabatic passage (STIRAP) is a fascinating physical process that dynamically explores chaotic and non-chaotic phases. In a recent paper Phys. Rev. Res. {\bf 2}, 042004 (R) (2020), such a phenomenon is realized in a cavity-QED platform. There, the emergence of chaos and its impact on STIRAP efficiency are mainly demonstrated in the semiclassical limit. In the present paper I treat the problem in a fully quantum many-body framework. With the aim of extracting quantum signatures of a classically chaotic system, it is shown that an out-of-time-ordered correlator (OTOC) measure precisely captures chaotic/non-chaotic features of the system. The prediction by OTOC is in precise matching with classical chaos quantified by Lyapunov exponent (LE). 
 Furthermore, it is shown that the quantum route corresponding to the semiclassical followed state encounters a dip in single-particle purity within the chaotic phase, depicting a consequence of chaos. A dynamics through the chaotic phase is associated with spreading of many-body quantum state and an irreversible increase in the number of participating adiabatic eigenstates.   
\end{abstract}

\maketitle
\section{Introduction}
Cavity-QED system is an interesting platform that holds immense potential of implementing quantum networks\cite{cirac,long,vogell,kato,meher,biswas}, quantum information processing and communication \cite{cirac,turchette,haroche,scully}, efficient quantum simulators for many-body systems  \cite{hartmann,mendoza,coto,jin} etc. The light-matter interaction in a cavity-QED offers Jaynes-Cumming (JC)-like \cite{tureci,ad,ad1} and Bose-Hubbard-like \cite{blais} nonlinearities at various regimes of parameters. Such nonlinear features result in exciting novel phenomena such as photon self-trapping \cite{tureci,ad,raftery}, nonlinear transport \cite{hughes,lahini}, and chaos \cite{ad1,larson,miguel,carlos}. 

Chaos, in classical systems, is well defined as the sensitivity to the initial condition and the exponential divergence of trajectories (with slightly different initial conditions) with time. However, the nature of manifestation, mechanism and diagnostic of chaos in the quantum counterpart of classically chaotic systems are relatively less established and is an active area of research. Equilibration of closed quantum systems is a fundamentally important open question and it can be a consequence of chaos in their classical counterparts\cite{eisert,lemos1}. Quantum chaos also delves into the deep connections among localization of quantum mechanical wavefunction, quantum-classical transition and decoherence mechanism\cite{zurek} and deals with some of the fundamental questions of physics. Therefore, understanding such features in complex quantum systems and its correspondence with classical counterpart demands a thorough theoretical and experimental investigation. A number of platforms have explored the quantum characteristics of classically chaotic systems. Some of them are optical realization of kicked harmonic oscillator\cite{lemos}, ultracold atoms \cite{christensen,moore,frisch}, atom-optics realization \cite{hainaut}, and cavity-QED setups \cite{carlos,swingle,zhu}.   

While the sensitivity to initial conditions for classical systems is quantified by LE, the corresponding quantum systems usually reflect integrability via level spacing statistics of eigenvalue spectrum \cite{wigner,izrailev}, participation number of eigenstates \cite{miguel,ad2}, OTOC \cite{larkin,hashimoto} etc. OTOC measures the dispersion of information (initially localized with a few degrees of freedom) to an exponentially large number of degrees of freedom, thereby resulting an apparent loss of local quantum information and distribution of correlation throughout the entire system \cite{preskill,shenker,maldacena}. This so called `scrambling' of quantum information is considered to be an efficient diagnostic of many-body quantum chaos \cite{hurtubise,swingle,zhu,swingle1,li,garttner,monroe}. Furthermore, OTOC is shown to be much reliable measure of quantum chaos compared to traditional level spacing statistics measure \cite{akutagawa}. The practical measurement of OTOC is quite challenging due to the need of back evolution during measurement \cite{hurtubise} and has been achieved in a limited number of systems \cite{li,garttner,monroe}. Therefore, seeking efficient strategies \cite{hurtubise} and physical systems with high-precision controllability are imperative. 

STIRAP is a process of remarkable utility and has been exploited in fields such as atomic population transfer, optical applications, state preparation and state transfer for quantum information processing and many more \cite{bergmann,vitanov}. The presence of `dark state' connecting only terminal nodes of a network, facilitates a robust adiabatic transfer which is immune to dissipation originating from intermediate nodes \cite{bergmann,vitanov}. Furthermore, a nonlinear STIRAP explores phases of varying integrability during a single sweep and is an excellent process to investigate chaos \cite{ad1,ad2,ad3,ad4}. The transition from non-chaotic (or regular) to chaotic phase simply by controlling tunable system parameter makes nonlinear STIRAP immensely interesting.  Efficient photon transfer protocol in a precisely controllable and scalable cavity-QED network is proposed in Ref.~\onlinecite{ad1}, where, chaos emerges during a STIRAP due to JC-like nonlinearity. In the present paper, I extend the analysis beyond the semiclassical treatment in Ref.~\onlinecite{ad1} and handle the problem in a quantum many-body framework. I show that the quantum process (corresponding to the semiclassical STIRAP) undergoes a spreading of its evolved many-body state within the chaotic parameter window, predicted by semiclassical theory. This is associated with irreversible increase in the participation number of adiabatic eigenstates of the system. A qualitative comparison is made between LE analysis (for semiclassical case) and microcanonical OTOC measure (for quantum many-body case). It shows a remarkable agreement in (semiclassically predicted) chaotic and non-chaotic regimes of the process. Additionally, the semiclassical followed state (which is nearly a `dark state') is shown to be constituted of a series of diabatic transitions through the avoided crossings between many-body eigenstates. The many-body state corresponding to this followed state produces a dip in single-particle purity within the chaotic window.

The paper is arranged as follows. In Sec. \ref{sec2} the model Hamiltonian is introduced and equations for stationary point (SP) solutions are deduced. Sec. \ref{sec3} deals with the quantum eigenspectrum and its features in the chaotic regime. LE analysis and OTOC measure are elaborated in Sec. \ref{sec4} and single-particle purity calculation for the quantum states (corresponding to the semiclassical followed state) is presented in Sec. \ref{sec5}. The chaotic effects on the slow-sweep real-time dynamics along with behavior of participation number is described in Sec. \ref{sec6}. Sec. \ref{sec7} deals with the sweep rate dependence of transfer efficiency and its behavior with varying photon number. Finally, we conclude and discuss potential future directions of our analysis in Sec. \ref{sec8}.

\section{Model, equations of motion, and scheme}\label{sec2}
The cavity-QED STIRAP Hamiltonian is given by
\begin{eqnarray}
 \hat{{H}}(\tilde{t})&=&\sum_{j\in \{a,b,c\}}\hat{H}_j - J_1(\tilde{t}) (\hat{a}^{\dagger}\hat{b}+h.c.) 
 \nonumber \\
 &&~~~~~~- J_2(\tilde{t}) (\hat{b}^{\dagger}\hat{c}+h.c.),
 \label{ham1}
\end{eqnarray}
where the JC Hamiltonian for cavity-a is defined as $\hat{H}_a=\omega_a \hat{a}^{\dagger}\hat{a}+\Omega_a \hat{s}^z_a+g_a(\hat{a}^{\dagger}\hat{s}^{-}_a+h.c.)$. $\hat{a}$ and $g_a$ are the photon destruction operator and JC coupling intensity for cavity-a, respectively. The qubit is described by the spin operator $s^{\alpha}_a$ ($\alpha \in \{x,y,z\}$). $\Omega_a$ and $\omega_a$ denote the frequencies for qubit and photon modes, respectively. The time-varying coupling parameters are $J_{1,2}(\tilde{t})=K{\rm exp}[-(\tilde{t}-\tilde{t}_{1,2})^2]$ (where the relation $\tilde{t}_1>\tilde{t}_2$ fixes the sequence of pump and Stoke's pulses) with the parametric time defined as $\tilde{t}\equiv t/\tau$ \cite{ad1}. This in turn parameterizes the time-dependent Hamiltonian $\hat{H}({t})$, making it only implicitly dependent on time $t$. The sweep rate of couplings is given by $\dot{\tilde{t}}=1/\tau$ and is a key factor for STIRAP dynamics \cite{ad1}. Throughout the paper we use the values $\tilde{t}_1=3.697$, $\tilde{t}_2=2.4242$ and consider the resonant case $\omega_{a,b,c}=\Omega_{a,b,c}$ for all the numerical results. To draw correspondence with the semiclassical results of Ref.~\onlinecite{ad1} we also consider a setup, where $g_a=g_b=0,g_c\neq 0$ and $\omega_{a,c}=\omega_b-\Delta$ (where $\Delta$ is central cavity detuning and is fixed at $\Delta=0.5K$ throughout the paper). Such a non-uniformity for $g_{a,b,c}$ has two aspects. Firstly, the Fock state $|n_a,n_b,n_c,s^z_c\rangle\equiv |N,0,0,-1/2\rangle$ becomes the eigenstate at $\tilde{t}=0$; this is advantageous for initialization keeping experimental realization in mind. Here, $n_{j},s_j^z$ are the eigenvalues of the operators $\hat{n}_j, \hat{s}^z_j$, respectively. Secondly, $g_b\neq0$ does not alter our findings regarding efficient adiabatic passage, because cavity-b remains negligibly occupied throughout the process. However, incorporating these assumptions and projecting $\hat{{H}}$ in the rotating frame of $\omega_a$ the new Hamiltonian can be rewritten as
\begin{eqnarray}
 \hat{H}(\tilde{t})&=&\Delta \hat{n}_b+g_c(\hat{c}^{\dagger}\hat{s}^{-}_c+h.c.) \nonumber \\
 &&- J_1(\tilde{t}) (\hat{a}^{\dagger}\hat{b}+h.c.) - J_2(\tilde{t}) (\hat{b}^{\dagger}\hat{c}+h.c.).
 \label{ham}
\end{eqnarray}
\begin{figure}[t]
  \centering
   \includegraphics[width=3.2in]{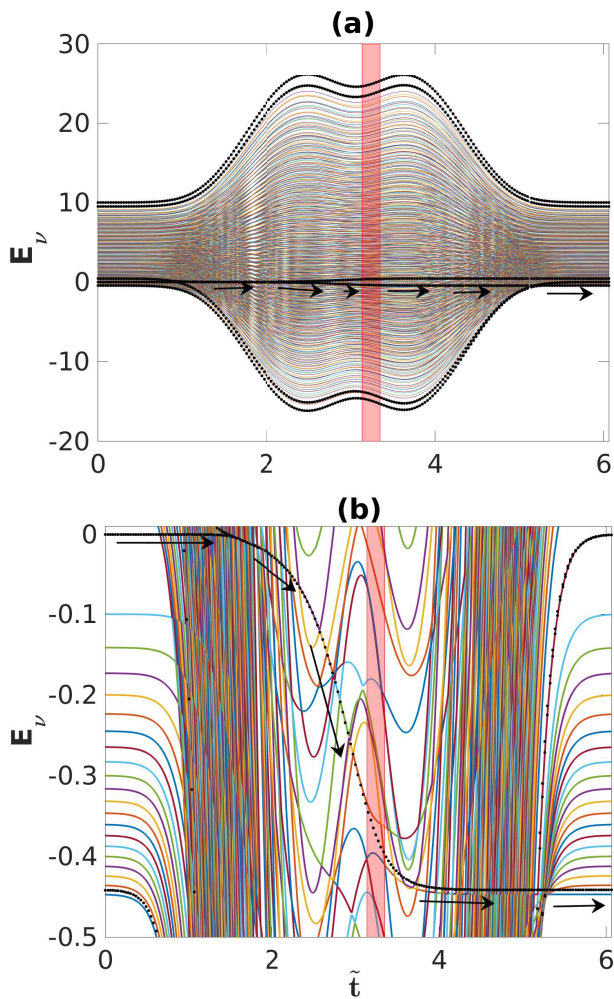}
  \caption{(Color online) For $g_c=0.1K$ quantum many-body eigenstates with varying $\tilde{t}$ are plotted as colored solid lines. The semiclassically obtained $E_{\rm SP}$ solutions are plotted by black dots and the red vertical patch marks the semiclassically obtained chaotic window of $\tilde{t}$, predicted by LE analysis in Sec. \ref{sec4}. (a) Full spectrum of many-body eigenstates are plotted. (b) Zoomed version of (a) showing that the followed SSP branch corresponds to a series of avoided crossings of quantum mechanical eigenstates. The black arrows in (a) and (b) direct the SSP branch $E_{ SSP}$ that leads to near-unity photon transfer from cavity-a to cavity-c with negligible occupancy of cavity-b.  }
  \label{fig1}
\end{figure}
The semiclassical approximation $\langle a^{\dagger} s^-_a \rangle \approx \langle a^{\dagger} \rangle \langle s^-_a\rangle$ is valid in large $N$ limit \cite{ad1,tureci}. We employ Heisenberg equation of motion with respect to $\mathcal{\hat{H}}(\tilde{t})=\hat{H}(\tilde{t})-\mu(\tilde{t})(\hat{n}_a+\hat{n}_b+\hat{n}_c+\hat{s}_c^z+1/2)$ and apply the above approximation. Furthermore, replacing the expectation values as $\{\langle \hat{a}\rangle,\langle \hat{b}\rangle,\langle \hat{c}\rangle, \langle \hat{s}^-_c\rangle,\langle \hat{s}^z_c \rangle\} \rightarrow \{a,b,c,s_c,s^z_c\}$ and setting the time derivatives to zero, we obtain the equations for stationary point (SP) solutions given by
\begin{eqnarray}
 J_1b+\mu a&=&0\label{sp1}, \\
 \Delta b-J_1 a-J_2 c-\mu b&=&0, \label{sp2}\\
 J_2 b-g_c s_c+\mu c&=&0, \label{sp3}\\  
 2g_c *c *s^z_c+\mu s_c&=&0. \label{sp4}
\end{eqnarray}
\begin{figure}[t]
  \centering
   \includegraphics[width=3.3in]{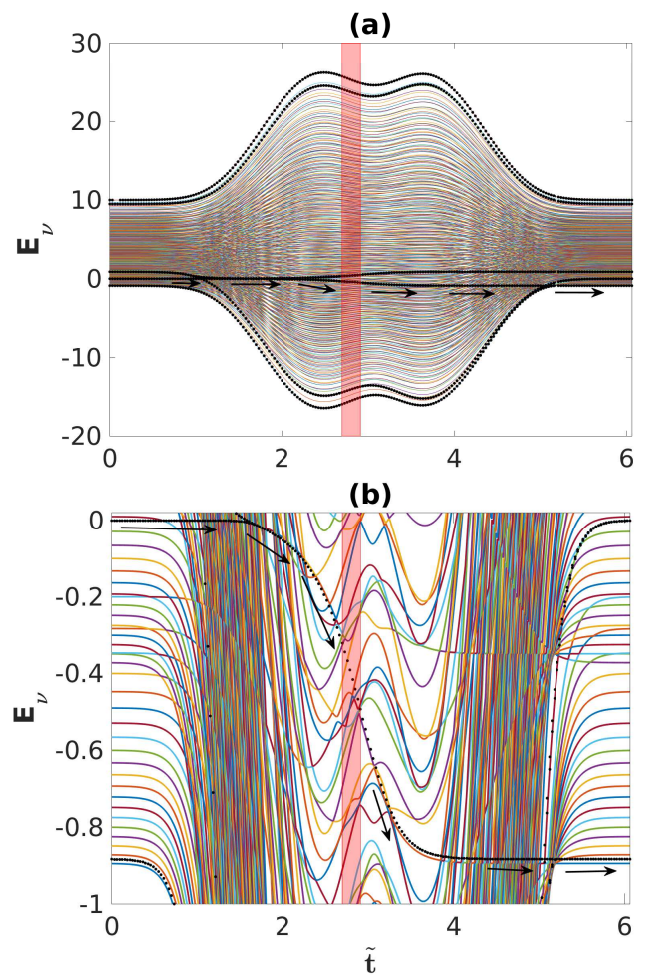}
  \caption{(Color online) For $g_c=0.2K$ quantum many-body eigenstates with varying $\tilde{t}$ are plotted as colored solid lines. The semiclassically obtained $E_{\rm SP}$ solutions are plotted by black dots and the red vertical patch marks the semiclassically obtained chaotic window of $\tilde{t}$, predicted by LE analysis in Sec. \ref{sec4}. (a) Full spectrum of many-body eigenstates are plotted. (b) Zoomed version of (a) showing that the followed SSP branch corresponds to a series of avoided crossings of quantum mechanical eigenstates. The black arrows in (a) and (b) direct the SSP branch $E_{\rm SSP}$ that leads to near-unity photon transfer from cavity-a to cavity-c with negligible occupancy of cavity-b.  }
  \label{fig2}
\end{figure}
Here $\mu$ is the chemical potential accounting for the conservation $\sum_{j\in \{a,b,c\}} n_j+[s^z_c+1/2]=N$ \cite{ad1}. The energies $E_{SP}$ corresponding to the SP solutions of Eqs.~ \ref{sp1}-\ref{sp4} are plotted as black dots in Figs.~\ref{fig1} and \ref{fig2}.  

The quantum dynamics of the system is dictated by Schr\"odinger's equation
\begin{eqnarray}
 i\dot{X}(t)=\hat{H}X(t),
 \label{schro}
\end{eqnarray}
where $X(t)$ is the quantum mechanical wavefunction, which is prepared as the Fock state $|X(0)\rangle=|N,0,0,-1/2\rangle$ at $t=0$ ($\tilde{t}=0$). For linear STIRAP (when $g_{a,b,c}=0$) the system state is prepared as $|X(0)\rangle=|N,0,0\rangle$ and the complete adiabatic sweep (from $\tilde{t}=0$ to $\tilde{t}=\tilde{t}_f$) translates the system to the desired state $|X(\tilde{t}_f)\rangle=|0,0,N\rangle$. This scheme is implementded by following the two-cavity coherent eigenstate \cite{ad2} given by
\begin{eqnarray}
 |\psi_{\rm d}\rangle_{\tilde{t}}=\frac{1}{\sqrt{N!}}[{\rm cos \Theta(\tilde{t})}\hat{a}^{\dagger}- {\rm sin \Theta(\tilde{t})}\hat{c}^{\dagger}]^N|{\rm vac}\rangle ,
 \label{coh}
\end{eqnarray}
where ${\rm cos}\Theta=J_2/\sqrt{J^2_1+J^2_2}$.
$|\psi_{\rm d}\rangle_{\tilde{t}}$ does not project on cavity-b and is equivalent to the semiclassical dark state. The nonlinear version of $|\psi_{\rm d}\rangle_{\tilde{t}}$ slightly deviates from Eq. \ref{coh} for moderate values of $g_c$. In the next section we elaborate the semiclassical adiabaticity when viewed from a quantum many-body perspective.

\section{Eigenspectrum and SP solutions} \label{sec3}
Here we plot the eigenvalue spectrum for various $g_c$ values and investigate its characteristics in the classically chaotic and non-chaotic regimes. The eigenenergies $E_{\nu}$ (with eigenstate index $\nu$) of the quantum Hamiltonian (Eq.~\ref{ham}) are obtained by diagonalizing $\hat{H}(\tilde{t})$ at various $\tilde{t}$'s and plotted by solid continuous lines in Figs.~\ref{fig1}, \ref{fig2}. In Fig. ~\ref{fig1} (a) $g_c=0.1K$ case is presented. There are $(N+1)^2$ (which is the size of Hilbert space) number of eigenvalues $E_{\nu}(\tilde{t})$ in contrast to only a few semiclassical $E_{SP}$'s. The particular $E_{SP}$ branch, that we follow to achieve cavity-a to cavity-c transfer by negligibly populating cavity-b, is designated by series of black arrows beside it. This {\it nearly dark} state (note, `dark state' is exactly defined for linear STIRAP $g_{a,b,c}=0$ \cite{ad1,vitanov}) is the special SP (SSP) branch of our interest. In Fig.~\ref{fig1} (b) it is clearly seen that the SSP branch ($E_{SSP}$) passes through a series of avoided crossings between many-body eigenstates. Therefore, the adiabatic following of classical SSP is actually numerous diabatic transitions among eigenstates (at avoided crossing location) in many-body scenario. Interestingly, unlike the avoided crossings in the non-chaotic region, the avoided crossings within the chaotic window of $\tilde{t}$ lack proximity between participating eigenstates. This actually portrays an enhancement of level repulsion within the chaotic window and disrupts the quantum diabatic route corresponding to the classical SSP branch. At the exit of the chaotic window the quantum route via {\it closely} avoided crossings is again formed along the SSP branch. Therefore, classically predicted chaos is remarkably manifested by enhanced level repulsion of quantum eigenspectrum, whereas the $E_{SSP}$ solutions gives no trace of chaos. It should be noted that the chaotic window (red patch in Figs.~\ref{fig1} (a) and (b)) is drawn by LE analysis presented in Sec. \ref{sec4}. To strengthen our observation we present $g_c=0.2K$ in Fig. ~\ref{fig2} and observe similar feature of eigenspectrum as in Fig. ~\ref{fig1}. Here too within the classically chaotic region the eigenvalues do not facilitate diabatic transitions by enhancing level repulsion. 

{At this point it is important to recall the standard energy level crossing prescription \cite{ad2} motivated by linear Landau-Zener transition\cite{landau,zener,stuckelberg,majorana}. For a pair of energy levels $\nu,\nu^{\prime}$ engaged in an avoided crossing, the diabatic transition between the levels requires $1/\tau\gg d^2_{\nu,\nu^{\prime}}/\sigma_{\nu,\nu^{\prime}}$, where $d_{\nu,\nu^{\prime}}=(E_\nu-E_{\nu^{\prime}})$ and $\sigma_{\nu,\nu^{\prime}}=|\langle \nu |\partial_{\tilde{t}}~\hat{H}(\tilde{t})|\nu^{\prime}\rangle|$ at the avoided crossing location. This depicts the dependence of dynamics on the rate of change of the Hamiltonian. In other words, when a pair of eigenvalues are widely apart and have smaller gradient at their closest proximity (which is the case for participating eigenlevels within the chaotic window), diabatic transition becomes challenging requiring very fast sweep of $J_{1,2}$.}

In the next section we quantify chaos for both semiclassical and quantum cases and draw a comparison.

\section{Lyapunov analysis and OTOC measure}\label{sec4}
\begin{figure}[t]
  \centering
   \includegraphics[width=3.5in]{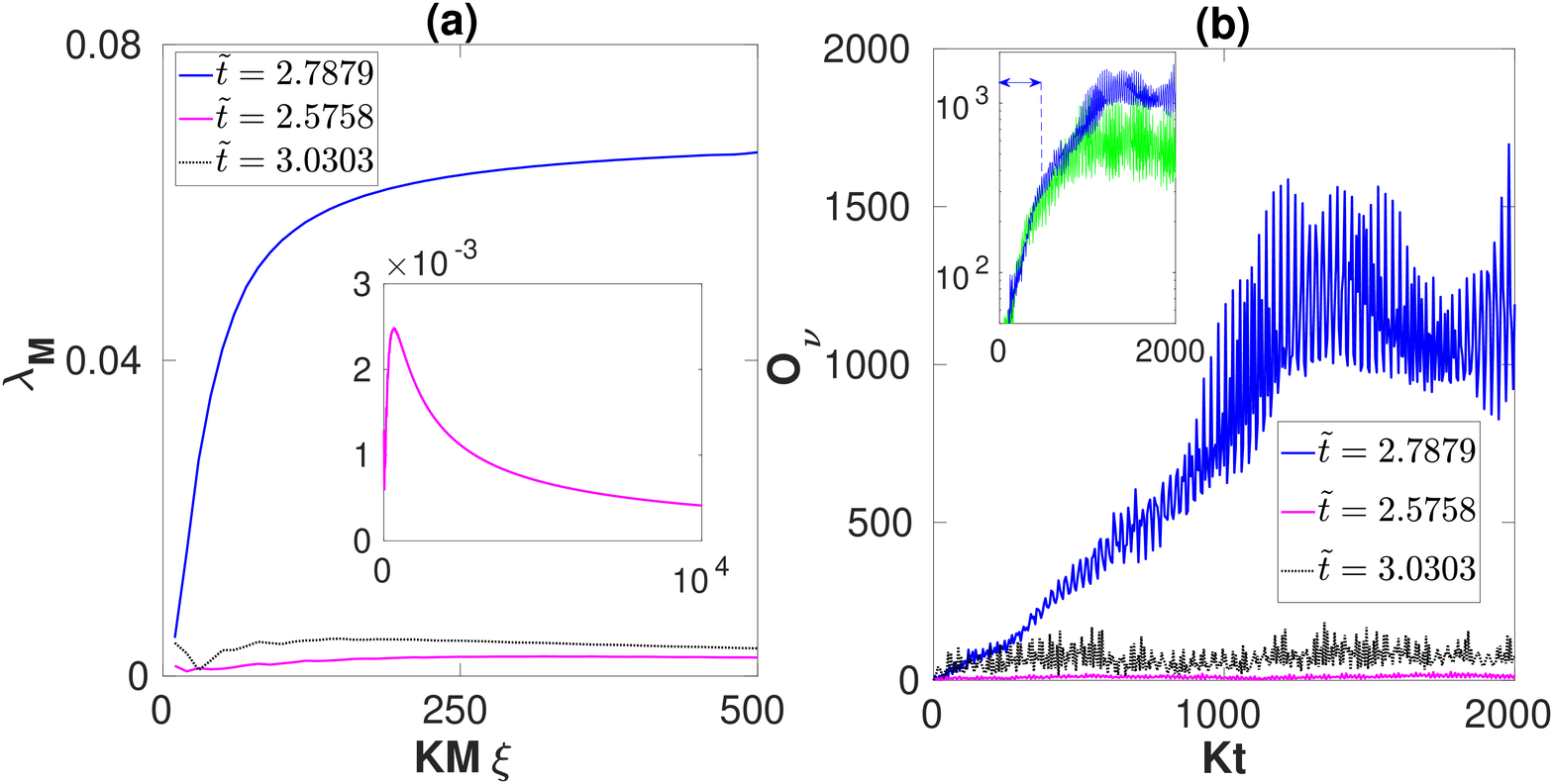}
  \caption{(Color online) Depicting quantum-classical comparision of chaos for $g_c=0.2K$. (a) LE plots at three representative $\tilde{t}$ values: $\tilde{t}=2.7879$ within the chaotic window introduced in Fig. \ref{fig2} and $\tilde{t}=2.5758, 3.0303$ at either sides of the window. (b) OTOC plots at same $\tilde{t}$ values for eigenstates (of the fixed Hamiltonian $H(\tilde{t})$) along semiclassical $E_{\rm SSP}[\tilde{t}]$. Here, the selective eigenstates are $|\nu\rangle\equiv |169\rangle, |164\rangle, |158\rangle$ for $\tilde{t}=2.5758, 2.7879, 3.0303$, respectively. The inset in (a) shows that the LE is asymptotically approaching zero for non-chaotic $\tilde{t}$. Inset in (b) plots log-scaled $O_\nu$ for the $\tilde{t}=2.7879$ case and an additional instance for $\tilde{t}=2.7273$ [green(grey) for $|\nu\rangle=|165\rangle$] within the chaotic window. Vertical blue dashed line marks exponential growth of $O_\nu$ followed by a saturation region.}
  \label{fig3}
\end{figure}

\begin{figure}[b]
  \centering
   \includegraphics[width=3.6in]{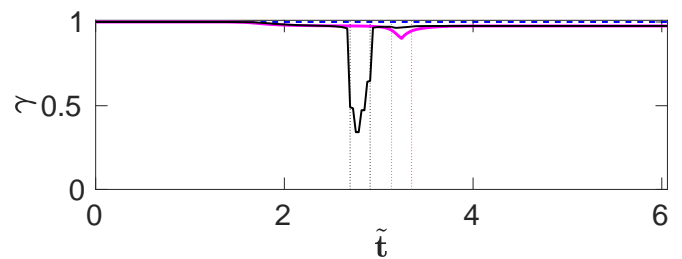}
  \caption{(Color online) One-particle purity at various $\tilde{t}$ for eigenstates along the SSP solutions. $g_c=0,0.1K,0.2K$ are plotted as dashed blue, solid magenta (grey), and solid black, respectively. The corresponding chaotic windows are marked by vertical dotted lines of same color.}
  \label{fig4}
\end{figure}

\begin{figure}[t]
  \centering
   \includegraphics[width=3.5in]{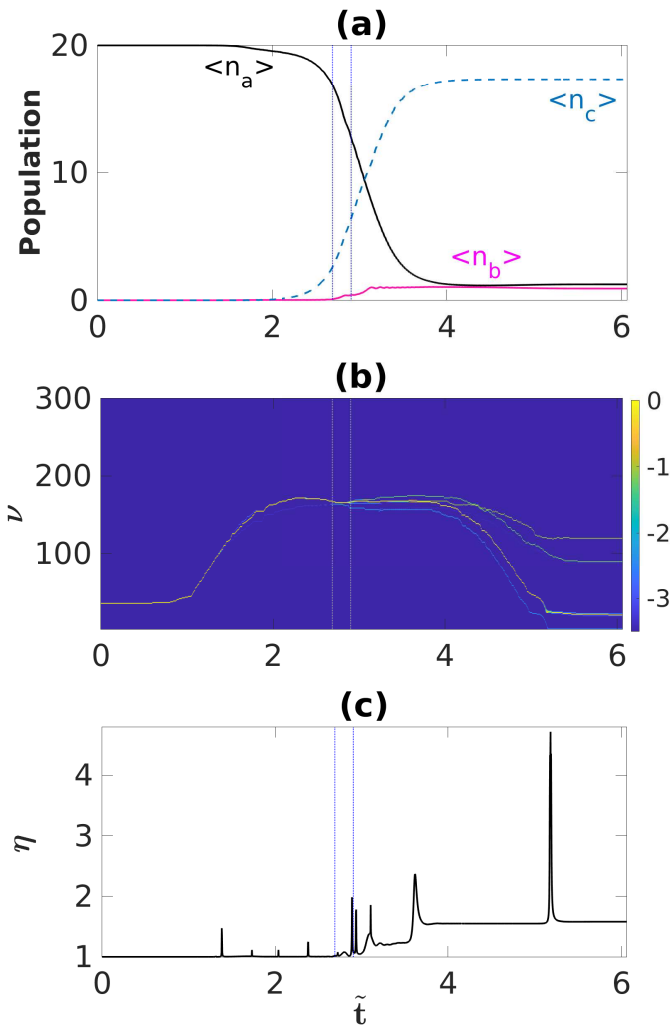}
  \caption{(Color online) (a) Real-time quantum dynamics of cavity populations for $g_c=0.2K$, when the system is initialized to $|n_a,n_b,n_c,s^z_c\rangle \equiv |20,0,0,-1/2\rangle$ at $\tilde{t}=0$ and swept at a rate $1/\tau=0.003K$. (b) Probability distribution over adiabatic eigenstates for dynamics described in (a). (c) Dynamics of participation number corresponding to (a). Vertical dotted lines in all three panels mark the chaotic window.}
  \label{fig5}
\end{figure}
For a classically chaotic system the phase space trajectories are extremely sensitive to the initial conditions. Two trajectories, which are infinitesimally separated in the phase space at $t=0$, exponentially diverge with time resulting completely varied outcome at long times. Chaos is quantified by the LE providing the rapidity of divergence. We consider a reference trajectory to be characterized by SSP solution at a particular $\tilde{t}$ and a test trajectory initially $\delta_0$ separated (w.r.t. reference trajectory) in the phase space. Both the trajectories are evolved for a time step $\xi$ followed by a reset of the test trajectory such that the new phase-space distance $\delta_1$ between the trajectories becomes $\delta_0$ along the direction same as $\delta_1$ \cite{scotti,Benettin,Benettin2,Benettin1,fine}. The procedure is repeated for a large number of steps $M$ and the LE is extracted as
\begin{eqnarray}
 \lambda_M=\lim_{\delta_0\to 0}\frac{1}{MK\xi}\sum^M_{j=1} {\rm log}\Big(\frac{\delta_j}{\delta_0}\Big).
 \label{le}
\end{eqnarray}
Here, $\delta_j$ denotes the phase-space distance just before the $j$-th reset.
The maximum LE $\lambda_{max}$ is defined for $M\to \infty$, whereas the finite-time LE $\lambda_M$ measures the divergence for a evolution time $t=KM\xi$. This procedure is advantageous compared to the usual procedure 
because the phase space for our system is bounded due to the constraint $n_a+n_b+n_c+s^z_c+1/2=N$ and this does not allow a monotonic growth of distance between the trajectories. 

Now for a quantum counterpart of the classically chaotic system chaos is captured by the exponential growth of commutator of observables, provided that the initial commutator value is considerably small \cite{hashimoto,preskill,shenker,maldacena,hurtubise,swingle,zhu,swingle1,li,garttner,monroe}. The thermal OTOC is defined as
\begin{eqnarray}
 O(t)_T&=&-\langle [U(t),V(0)]^2\rangle_T,
 \label{otoc}
\end{eqnarray}
where $U$ and $V$ are two operators; $\langle...\rangle_T$ denotes the thermal average defined as $\langle \hat{A} \rangle_T=Z^{-1}\sum_n \langle n| \hat{A} |n\rangle.e^{-\beta\epsilon_n}$, where $\beta=1/k_BT$, $\epsilon_n$ is the energy of the energy eigenstate $|n\rangle$, and $Z$ is the partition function. In our analysis we only focus on the particular energy eigenstates corresponding to the followed SSP branch and deal with the microcanonical OTOC defined as $O_\nu(t)=\langle \nu |[\hat{n}_a(t),\hat{n}_c(0)]^2|\nu\rangle$ \cite{hashimoto,carlos}. At $t=0$ $\hat{n}_a$ and $\hat{n}_c$ commute. Expanding $O_\nu(t)$ in the energy eigenbasis we write it as 
\begin{eqnarray}
 O_{\nu}(t)=\sum_{\nu^{\prime}}~\Big| \langle \nu | [\hat{n}_a(t),\hat{n}_c(0)] |\nu^{\prime}\rangle \Big|^2.
 \label{otoc_micro}
\end{eqnarray}
In Fig. \ref{fig3} (a) we plot the LE (corresponding to $E_{SSP}$) at three representative $\tilde{t}$ values and obtain a positive LE for $\tilde{t}=2.7879$ falling within the chaotic window marked in Fig. \ref{fig2}. On the contrary, $\lambda_{max}\rightarrow 0$ for $\tilde{t}$ on either side of the window, indicating non-chaotic regions \cite{ad1}. Fig. \ref{fig3} (b) plots the microcanonical OTOC for eigenstates having energies closest to $E_{SSP}$. We observe an exponential increase in $O_\nu$ (inset of Fig. \ref{fig3}(b)) only for $\tilde{t}$ that produces a positive $\lambda_{max}$ in its semiclassical counterpart. The exponential growth of $O_{\nu}$ tends to saturate at long times. 

\section{One-particle purity}\label{sec5}
In this section we analyze the single-particle purity along the quantum mechanical diabatic route for cavity-a to cavity-c transfer of photons. The single-particle purity is defined as
\begin{eqnarray}
 \gamma={\rm Trace}([\rho^{sp}]^2).
 \label{purity}
\end{eqnarray}
Here the single-particle reduced density matrix is defined as
\begin{eqnarray}
\rho^{sp}_{i,j}=(1/N)\langle \nu|\hat{A}_i^{\dagger}\hat{A}_j|\nu\rangle, 
\end{eqnarray}
where $\hat{A}_i\equiv\{\hat{a},\hat{b},\hat{c},\hat{s}^-_c\}$ and $|\nu\rangle$ is the eigenstate of interest. Note that the flipping of qubit from ground to excited state is associated with absorption of one photon and $\hat{s}^+_c\hat{s}^-_c$ is quantitatively equivalent to one photon excitation. $\gamma$ falls within the range $\{1/4,1\}$. When $\gamma=1$ the many-body eigenstate is a coherent state having localized distribution in the phase space. On the other hand, $\gamma=1/4$ corresponds to a maximally mixed state. In Fig. \ref{fig4} we plot $\gamma$ along the quantum diabatic route corresponding to the classical adiabatic SSP branch. We pick $|X_\nu\rangle$ along the classically adiabatic route $E_{SSP}$ and calculate $\gamma$. We observe that $\gamma$ has a sharp dip in the regime where the system is classically chaotic. Moreover, for stronger $g_c$ the depth of the dip increases. This is because the phase space distribution of the adiabatic eigenstate spreads over the chaotic zone and considerably deviate from being a coherent state \cite{ad2}. Therefore, the strong reduction of single-particle purity is reflected here as a consequence of chaos.

So far we have dealt with features of eigenspectrum at various $\tilde{t}$'s falling inside and outside the chaotic window. In the subsequent sections we explore their consequences on the real-time dynamics of the system and on the STIRAP efficiency.  
\section{Quantum dynamics and chaotic spreading}\label{sec6}
Here we investigate the chaotic effects on the real-time dynamics of the system. It is to be noted that the dynamics very much depends on the sweep rate $1/\tau$ \cite{ad1} (i.e., how fast the system Hamiltonian is tuned), although the eigenspectrum only depends on the parametric time $\tilde{t}$ (i.e., on $\hat{H}(\tilde{t})$ in Eq.~\ref{ham}). The slower-sweep semiclassical dynamics initiated at SSP branch at $t=0$ starts oscillating about the SSP branch within the chaotic window, thereby diminishing the transfer efficiency \cite{ad1}. For a quantum case we initiate the system in the Fock state $|X(0)\rangle \equiv |N,0,0,-1/2\rangle$ and the dynamics is studied using Schr\"odinger equation given by Eq. \ref{schro}. Within the chaotic window the evolved many-body state $|X(t)\rangle$ suffers a spreading over the adiabatic eigenstates supported by the chaotic energy range. 

In Fig. \ref{fig5} (b) we plot the probability distribution over the adiabatic eigenstates while the slow-sweep dynamics is carried out. It is clear that the system's many-body state spreads out exactly within the semiclassically predicted chaotic window and the consequence is reflected in the diminished transfer (from cavity-a to cavity-c) in Fig. \ref{fig5} (a). Moreover, the cavity-b population starts increasing within the chaotic window deteriorating the STIRAP protocol. The spreading can be quantified by the number of eigenstates participating in the evolved state $|X(t)\rangle$ and is given by 
\begin{eqnarray}
 \eta(t)=1/\sum_\nu |\langle \nu | X(t) \rangle|^4.
 \label{pn}
\end{eqnarray}
In Fig. \ref{fig4}(c) $\eta$ shows only reversible spikes ($1<\eta<2$) before reaching the chaotic window. These occur due to the diabatic transitions while evolving. In contrast, the increment of $\eta$ within the chaotic window irreversibly goes beyond 2 and distinctively portrays a non-chaotic-to-chaotic transition. Therefore, the signature of chaos in a quantum framework is expressed as spreading of the evolved state and irreversible growth of participation number \cite{ad2,miguel}. 

In the next section we explore transfer efficiency with varying sweep rate and investigate the physics when the total excitation number is varied.
\begin{figure}[t]
  \centering
   \includegraphics[width=3.5in]{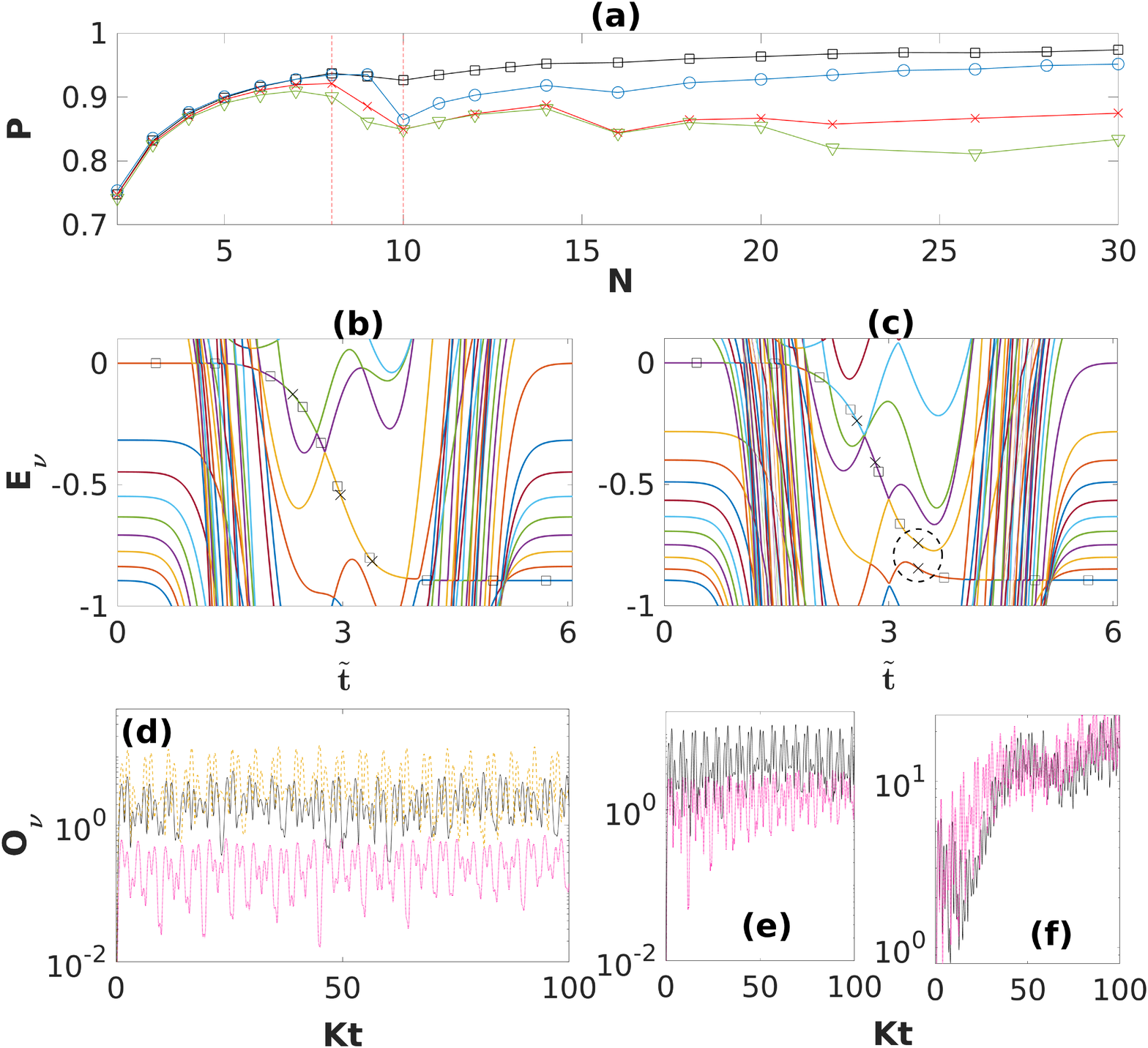}
  \caption{(Color online) (a) Efficiency $P$ plotted with varying total excitation number $N$ and for $g_c\sqrt{N}=0.8944K$, $\Delta N=10K$, $J_{1,2}N=20K{\rm exp}[-(\tilde{t}-\tilde{t}_{1,2})^2]$. Cases with sweep rates $1/\tau=0.0606K,0.0152K,0.003K,0.0015K$ are plotted by square, circle, `$\times$' and triangle line types, respectively. Vertical dashed lines mark $N=8$ and $10$. The eigenvalue spectrums are plotted for (b) $N=8$ and (c) $10$, respectively. Squares in (b) and (c) describe the route (composed of series of avoided crossings) that leads to cavity-a to cavity-c photon transfer. The dashed circle in (c) indicates the region where the energy gap between participating eigenstates becomes much wider, disrupting the diabatic crossing. (d), (e), (f) $O_{\nu}$ is plotted in logarithmic scale for eigenstates on the followed route at various $\tilde{t}$ and these states are marked by `$\times$' marks in (b) and (c). $O_{\nu}$ plots in (d) corresponding to (b) are for $\{\tilde{t},\nu\}:$ $\{2.3333,33\}$ (dotted magenta), $\{2.9697,31\}$ (dashed orange), and $\{3.3939,31\}$ (solid black). (e) corresponds to (c) for $\{2.5758,48\}$ (dotted magenta), $\{2.8182,46\}$ (solid black). (f) corresponds to (c)  for $\{3.3939,44\}$ (dotted magenta) and $\{3.3939,45\}$ (solid black).}
  \label{fig6}
\end{figure}
\section{Sweep-rate dependence and photon number dependence} \label{sec7}
So far we have seen that the quantum signature of chaos is consistent with the semiclassical prediction for relatively larger values of $N$. Here we investigate the chaotic features when the total excitation $N$ is varied. In Fig. \ref{fig6} (a) we plot the efficiency $P=\langle \hat{n}_c\rangle_{end}/N$, where $\langle \hat{n}_c \rangle_{end}$ is the expectation value of cavity-c population at the end of STIRAP scheme. The reduction in efficiency with slower sweep rates describes the `slower is worse' behavior in the presence of chaos \cite{ad1,ad3}. Such an outcome originates from the fact that a slower sweep permits the system spending longer time within the chaotic window and enhance the spreading of evolving state (see Fig. \ref{fig5}). Another aspect of Fig. \ref{fig6} (a) is the appearance of chaotic features with the varying photon number $N$. 
In Fig. \ref{fig6}(a) $P$ is plotted for various $N$ by keeping the characteristic parameters $J_{1,2}N$, $\Delta N$, and $g_c\sqrt{N}$ fixed \cite{tureci}.
We observe that up to some $N$ (e.g., $N=8$ for $1/\tau=0.003K$) the efficiency smoothly increases for a particular $1/\tau$. This is due to the fact that, with increasing $N$ $P$ becomes larger compared to the excitation shared with the qubit in cavity-c. The cavity-c state entangles photon and qubit degrees of freedom. Following this there is a sudden decrease in efficiency with further increment of $N$ and a non-monotonic jagged behavior thereafter. $P$ in this regime of $N$ considerably varies for various $1/\tau$ values and decreases with smaller $1/\tau$. Therefore, emergence of chaos demands certain level of complexity in the quantum system and the complexity results from larger photon number. To clarify this point we plot Fig. \ref{fig6} (b), (c), where the {\it widely}-avoided-crossing feature (as pointed out in Sec. \ref{sec3} and marked by circle in Fig. \ref{fig6} (c)) can only be seen in Fig. \ref{fig6} (c) for $N=10$. This feature in the eigenspectrum nicely correlates with the onset of chaotic (jagged) behavior in Fig. \ref{fig6} (a). Furthermore, OTOC plots in Figs. \ref{fig6} (d), (e), and (f) confirm chaotic eigenstates in the encircled region of Fig. \ref{fig6} (c).  Another observation is that the efficiency for faster sweep rates monotonically increases for larger $N$, whereas for slower rates its behavior is jagged. This hints that chaotic disruption takes place at slower sweep rates, whereas a sufficiently faster sweep dodges such effects. 

\section{Conclusion and discussion}\label{sec8}
In this paper we have investigated the quantum signatures of chaos in a c-QED based STIRAP. Throughout the paper we deal with a Hermitian case but in a realistic system non-Hermitian contributions are activated through cavity loss and qubit decay. However, our analysis can well be justified when the decay rates are considerably smaller than the sweep rate [see supplementary material of \cite{ad1}]. In this situation the STIRAP scheme is completed well before the dissipation affects significantly. 

Our analysis is demonstrated in an experimentally realizable setup of c-QED STIRAP.
Well developed techniques of high-precision state preparation, measurement, and control of the cavity-QED system, make c-QED STIRAP an interesting platform for testing quantum chaos signatures and design protocols for efficient population transfer in nonlinear realistic systems. As a future direction, it is interesting to extend such protocols in larger c-QED networks and investigate steady-state properties in driven-dissipative version of the studied case.

\end{document}